\documentclass[usenatbib]{mnras}

\usepackage{newtxtext,newtxmath}
\usepackage[T1]{fontenc}
\usepackage{graphicx}	% Including figure files
\usepackage{amsmath}	% Advanced maths commands
\title{Probing the kinematics and chemistry of the hot core Mon R~2 IRS~3 using ALMA observations}
\author[A. Fuente et al.]{
A. Fuente,$^{1}$\thanks{E-mail: a.fuente@oan.es}
S. P. Trevi\~no-Morales$^{2}$,
T. Alonso-Albi$^1$,
A.~S\'anchez-Monge$^3$, 
\newauthor
P. Rivi\`ere-Marichalar$^{1}$,
and D. Navarro-Almaida$^1$
\\
$^{1}$Observatorio Astron\'omico Nacional (OAN,IGN), Alfonso XII, 3,  28014, Madrid, Spain\\
$^{2}$Chalmers University of Technology, Department of Space, Earth and Environment, SE-412 93 Gothenburg, Sweden\\
$^{3}$Harvard-Smithsonian Center for Astrophysics, 60 Garden St., Cambridge, MA 02138, USA\\
$^{4}$Institut de Recherche en Astrophysique et Planétologie, 9 avenue du colonel Roche, 31028 Toulouse Cedex 4, France\\
}

\date{Accepted 2021 July 14. Received 2021 July 14; in original form 2021 March 8}

\pubyear{2021}

\begin{document}
\label{firstpage}
\pagerange{\pageref{firstpage}--\pageref{lastpage}}
\maketitle


\begin{abstract}
We present high angular resolution 1.1mm continuum and spectroscopic ALMA observations of the well-known massive proto-cluster Mon~R~2~IRS~3.The continuum image at 1.1mm shows two components, IRS~3~A and  IRS~3~B, that are separated by $\sim$0.65$"$. We estimate that IRS~3~A is responsible of $\sim$80 \% of the continuum flux, being the most massive component. We explore the chemistry of IRS~3~A based on the spectroscopic observations. In particular, we have detected intense lines of  S-bearing 
species such as SO, SO$_2$, H$_2$CS and OCS, and of the Complex Organic Molecules (COMs) methyl formate (CH$_3$OCHO) and dimethyl ether (CH$_3$OCH$_3$). The integrated intensity maps of most  species show a compact clump centered on IRS~3~A,  except the emission of the
COMs that is more intense towards the near-IR nebula located to the south of IRS~3~A, and HC$_3$N whose emission peak is located  $\sim$0.5$"$ NE from  IRS~3~A. The kinematical study suggests that the molecular emission is mainly coming from a rotating ring and/or an unresolved disk.
Additional components are traced by the ro-vibrational HCN $\nu_2$=1 3$\rightarrow$2 line which is probing the inner disk/jet region,
and the weak lines of CH$_3$OCHO,  more likely arising from the walls of the cavity excavated by the molecular outflow.
Based on SO$_2$  we derive a gas kinetic temperature of T$_k$$\sim$ 170 K towards the IRS~3~A.  The most abundant S-bearing species is SO$_2$ with an abundance of $\sim$ 1.3$\times$10$^{-7}$, and $\chi$(SO/SO$_2$) $\sim$ 0.29. Assuming the solar abundance, SO$_2$ accounts for $\sim$1\% of the sulphur budget.
\end{abstract}


%%%%%%%%%%%%%%%%%%%%%%%%%%%%%%%%%%%%%%%%%%%%%%%%%%

%%%%%%%%%%%%%%%%% BODY OF PAPER %%%%%%%%%%%%%%%%%%

\section{Introduction}
\label{sec:intro}
In spite of its importance for galaxy evolution, the formation of massive stars is poorly known yet (for reviews, see, e.g., \citealp{Beuther2007, 
Zinnecker2007,Smith2009,Tan2014,Motte2018}). High-mass star forming
regions are less abundant than their low-mass counterparts and are typically located at large distances ($>$ 1 kpc), 
making the study of individual collapsing objects challenging. Furthermore, massive stars form in tight clusters that 
cannot be resolved with single-dish observations. In spite of these difficulties, a big progress has been done in
recent years based mainly on observational studies, and the high mass star forming regions are
now divided into several evolutionary stages \citep{Beuther2007,Zinnecker2007}. The formation of massive stars begins in massive clouds that are 
seen as infrared dark clouds (IRDCs) when placed in front of bright infrared emission. These massive clouds that harbor 
high-mass starless cores and low- to intermediate-mass protostars (e.g., \citealp{Pillai2006, Rathborne2006}) evolve to form High Mass Protostellar Objects (HMPOs) with  
$>$8~M$_\odot$, showing gas accretion and molecular outflows (e.g., \citealp{Beuther2002, Motte2007}). The massive protostars warm the surrounding material to temperatures higher than 100~K. This is the hot core phase  during which the icy mantles
of dust grains evaporate, chemically enriching the gas with S-bearing and complex molecules (see e.g., \citealp{Charnley1997, Viti2004, 
Wakelam2004, Garrod2008, Herbst2009, Herpin2009}). Numerous millimeter and submillimeter wavelength observations
towards hot cores reveal a rich chemistry in nitrogenated and oxygen-bearing complex molecules (COMs) (for a review, see \citealp{Herbst2009}).  
However, many of them are based on single-dish observations that do not allow to resolve the individual cluster 
 members and disentangle the physical and chemical complexity. High spatial resolution observations of
hot cores are required to investigate this important  stage in the formation of massive stars. In this paper, we use 
ALMA observations to probe the physical and chemical structure of the Mon R~2 IRS~3 hot core, that is located in the Monoceros R2 giant molecular cloud.

\begin{figure*}
    \centering
     \includegraphics[width=1.0\textwidth]{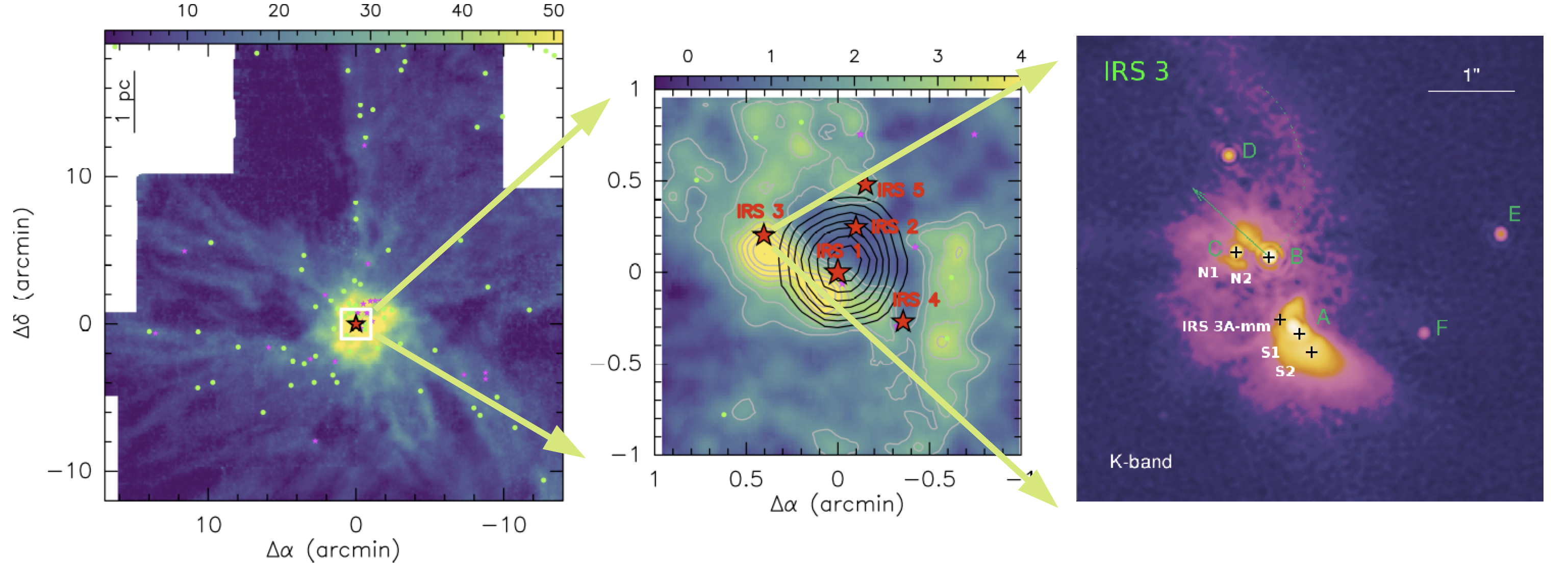}
    \caption{{\it Left:} Hub-filament system in Mon~R2 as seen with the $^{13}$CO (1--0) molecular emission \citep{Trevino2019}. The white box corresponds to the inner 0.7 pc $\times0.7$~pc around the cluster and zoomed in the right panel. {\it Centre}:  Central hub of Mon~R2 as seen with the H$^{13}$CO$^+$ (3→2) molecular emission (in colours and gray contours) tracing the high-density molecular gas \citep{Trevino2014}. The black contours corresponds to the H$\alpha41$ emission tracing the HII~region \citep{Trevino2019}. The red star gives the positions of the IRS 1-5  infrared sources.The colored symbols are the sources identified by \citet{Rayner2017}: pink stars are protostars, and the green circles are bound clumps. {\it Right:} Color representation of the K band speckle image reported by \citet{Preibisch2002}. Green letters correspond to the NIR sources as reported by  \citet{Preibisch2002}. Black crosses with white labels correspond to the positions listed in Table~\ref{table:points} that we have defined for
    an easier comparison with observations.}
    \label{fig:positions}
\end{figure*}

Monoceros~R2 (hereafter Mon~R2) is an active massive star-forming cloud located at a distance of $\sim$~778~pc \citep{Zucker2019}. 
Recently, several works \citep{Didelon2015, Pokhrel2016, Rayner2017, Trevino2019} have revealed an intriguing look of the cloud. The properties of 
Mon~R2 are in agreement with a scenario of global non-isotropic collapse where  the gas flows along several filaments that  converge
into the central hub ($\sim$ 2.25~pc$^{2}$; see Fig~\ref{fig:positions} and \citealp{Trevino2019}), feeding it with an accretion rate of $10^{-4}-10^{-3}$ M$_\odot$~yr$^{-1}$. The filaments extend within the central hub forming a complex structure that shows signs of rotation and infall motions, with the gas falling into the stellar cluster (see \citealp{Trevino2019}).  The cluster contains hundreds of stars which are obscured by an average visual extinction of 33 mag \citep{Carpenter1997}. 
The brightest infrared sources are IRS~1, IRS~2, IRS~3, IRS~4 and  IRS~5
with  luminosities of $\sim$3000~L$_\odot$, $\sim$6500~L$_\odot$, $\sim$14000~L$_\odot$, $\sim$700~L$_\odot$, 
and  $\sim$300 L$\odot$, respectively  \citep{Henning1992, Hackwell1982}.
Among these, the most massive star is IRS~1, at RA(J2000) = $06^{h}07^{m}46.2^{s}$, Dec(J2000) = $-06^\circ23'08.30''$, with a mass of $\sim$12 M$_\odot$
(e. g., \citealp{Thronson1980, Giannakopoulou1997}). This source is driving an expanding ultracompact (UC) HII region creating a cavity free of molecular 
gas extending for about 0.12 pc \citep{Choi2000, Dierickx2015} and surrounded by a number of photon-dominated regions (PDRs) with different 
physical and chemical conditions (e. g., \citealp{Ginard2012, Pilleri2012, Pilleri2013, Trevino2014, Trevino2016}). 
IRS~2 is very compact and does not show any structure at sub-arcsecond scales 
\citep{Alvarez2004,Jimenez2013}. IRS~3 and IRS~5 have not any associated HII region, and are invisible at optical wavelengths, 
consistent with being in an earlier evolutionary  stage.  \citet{Dierickx2015} reported interferometric images of the Mon~R2 cluster in millimeter 
continuum and molecular line emission carried out
with the Submillimeter Array (SMA) at angular resolutions ranging from 0.5$"$ to $\sim$3$"$. The detection of 
molecular tracers such as CH$_3$CN, CH$_3$OH or SO$_2$ towards IRS~3 and IRS~5
confirmed that these sources are massive young stellar objects. Moreover, the gas temperatures derived
from these observations, $>$ 100~K, probe that they are in the hot core stage.

Because of its youth, high luminosity, and location close to the Sun,
Mon~R2~IRS~3 is a well known hot core that has been observed at essentially all wavelengths.
Sub-arcsecond infrared imaging by \citet{Beckwith1976} indicated that IRS~3 is a double source.
This was later confirmed by \citet{McCarthy1982}, who derived a separation of 0.87$"$ at a position angle of 13.5$^\circ$.
\citet{Koresko1993} carried out speckle interferometric imaging of IRS~3 in the near infrared (NIR)
K(2.2~$\mu$m) and L$'$ (3.8~$\mu$m) bands and at 4.8$\mu$m. This study revealed a bright conical nebula 
associated to the southern component, and a previously unknown compact source 0.37$"$ east of the northern component. 
Further NIR speckle observations by  \citet{Preibisch2002} showed that IRS~3 is in fact a cluster of at least 6 NIR sources  (see Table~\ref{table:points}),
one of which, B, shows a microjet (see Fig.~\ref{fig:positions}) located at a position angle, P.A.=50$^\circ$  
from north to east. A high-velocity CO molecular outflow (v$_{out}$ $\sim$ 30~km~s$^{-1}$) was later detected 
at a similar direction by \citet{Dierickx2015}. However, the spatial resolution of these observations  do  not allow to discern whether the outflow is driven
by the NIR source A or B, and  the fan-like structure of the blue-shifted low introduces  some uncertainty in the determination of the outflow  direction. 
Higher spatial resolution interferometric observations using the MIDI instrument of the Very Large
Telescope Interferometer (VLTI)  in the N band (8--13~$\mu$m) detected 
compact emission towards NIR sourced A and B \citep{Boley2013}. The most intense component at 10~$\mu$m
was detected towards A with an emission size of  FWHM$\sim$164 mas. The emission towards 
B was more compact, FWHM$\sim$38 mas. 
%The analysis of the correlated
%fluxes suggests that the disk around IRS~3B is at low inclination, close to the
%face-on morphology.

All the mm studies carried out. thus far towards IRS~3 were performed with single-dish observations or interferometric observations which are unable to resolve the  IRS~3 mini-cluster
(see \citep{Boonmanvd2003, vandertak2003, Dierickx2015}.
Recently, \citet{Dungee2018} observed the $\nu_3$ ro-vibrational band of SO$_2$ confirming the existence of hot molecular gas (T$_k$=234$\pm$15 K) 
towards IRS~3, and derived a high SO$_2$abundance ($\sim$ a few 10$^{-7}$).  Based on high spatial resolution continuum and
spectroscopic ALMA observations,  in the following we investigate the kinematics (Sect.~\ref{sec:mom}) and chemistry (Sect.~\ref{sec:mol}) of this massive protostar
that is a reference for massive star formation studies.

\begin{table}
	\caption{Positions.}
	\label{table:points} 
	\centering                      
	\begin{tabular}{lll}       
	\hline\hline
Position       &RA     &DEC \\ \hline
N1   &  06:07:47.880  &  $-$06:22:55.42 \\
N2   &  06:07:47.855  &  $-$06:22:55.49 \\
IRS~3 A - mm    &   06:07:47.847  &  $-$06:22:56.21 \\   
%C    &  06:07:47.840  &  $-$06:22:56.11 \\
S1       &  06:07:47.833  &  $-$06:22:56.36 \\
S2       &  06:07:47.822  &  $-$06:22:56.58 \\
%NIR 06:07:47.836 -06:22:56.29
%
\hline                                   
\end{tabular}
\end{table}
%
% Offset N2:  +0.22 , +0.62
%
%
\begin{figure*}
    \centering
    \includegraphics[width=1.0\textwidth]{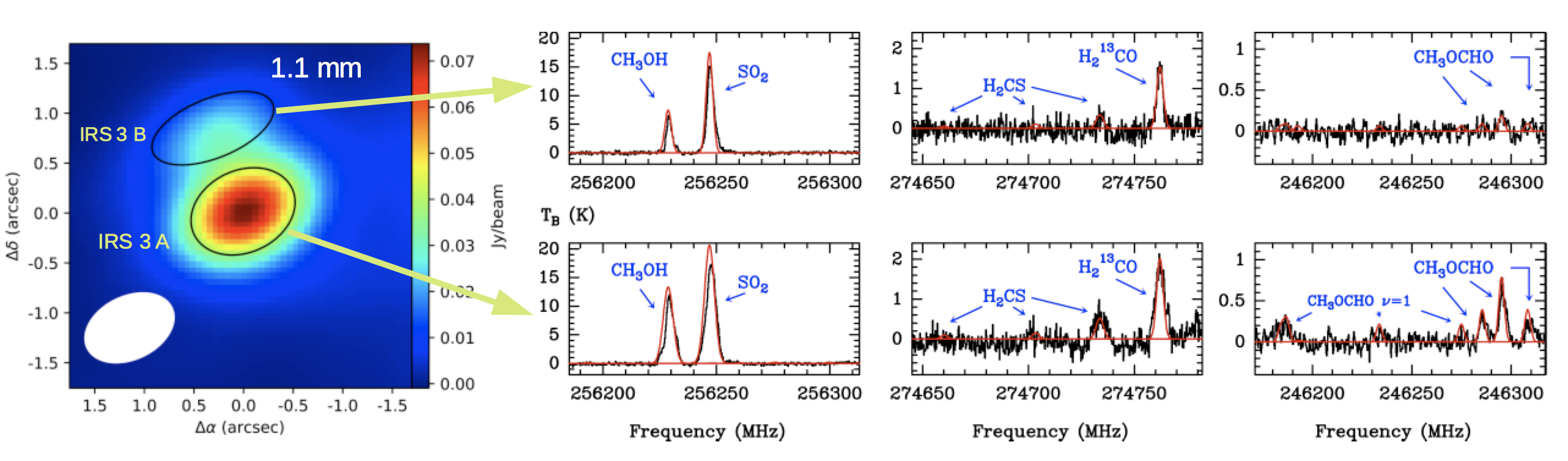}
    \caption{{\it Left:} Continuum image created by merging the bands centered  at 262~GHz and 272~GHz  (1.1 mm). The rms of the image is $\sigma$=3 mJy/beam. The black ellipses indicate the  Half Power Beam Width of the Gaussians fitted to IRS~3~A (south) and IRS~3~B (north), respectively.
    {\it Right)} Example of the spectra observed towards  IRS~3~A  and IRS~3~B.  .
    }
\label{fig:continuum}
\end{figure*}

\begin{figure*}
    \centering
    \includegraphics[width=1.0\textwidth]{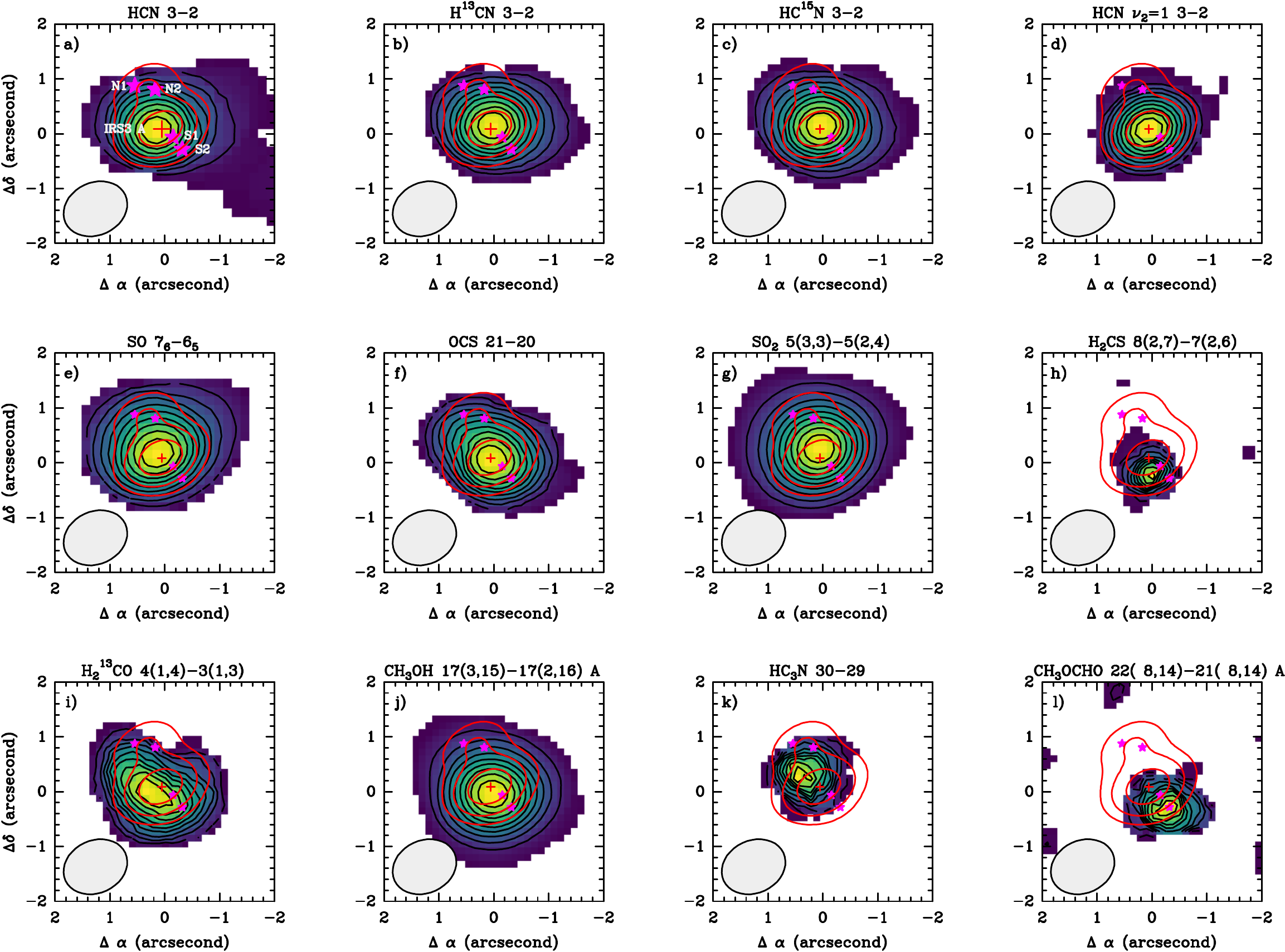}
    \caption{Line Integrated intensity maps  calculated in the 0 to 20 km s$^{-1}$ velocity  interval for all the lines except the HC$_3$N 30$\rightarrow$29 and CH$_3$OCHO 22(8,14)$\rightarrow$21(8,14) lines for which a narrower velocity range, from 5 to 15  km s$^{-1}$, was selected. All maps have been produced using an intensity threshold of 3$\sigma$.
Contours (in black) correspond to 10\% to 90\% of the peak emission in steps of 10\%. First contour and step are:  a) 35.00~K~km~s$^{-1}$;  b) 11.204~K~km~s$^{-1}$; c) 4.149~K~km~s$^{-1}$;  d) 3.048~K~km~s$^{-1}$; e) 33.031~K~km~s$^{-1}$; f) 2.499~K~km~s$^{-1}$; g) 11.272~K~km~s$^{-1}$; h) 0.224~K~km~s$^{-1}$; i) 0.824~K~km~s$^{-1}$; j) 6.938~K~km~s$^{-1}$; k) 0.251~K~km~s$^{-1}$; l) 0.174~K~km~s$^{-1}$. The beam is shown in the left-bottom. corner. We have superposed in red contours the continuum emission at 1.1~mm (levels are 15, 30, 60, 90 mJy/beam). Stars indicate the positions N2, N1, S1, and S2 in Table~\ref{table:points}. The cross indicates the position of the millimeter source IRS~3~A.
}
\label{fig:maps}
\end{figure*}

\begin{figure}
    \centering
    \includegraphics[width=0.4\textwidth]{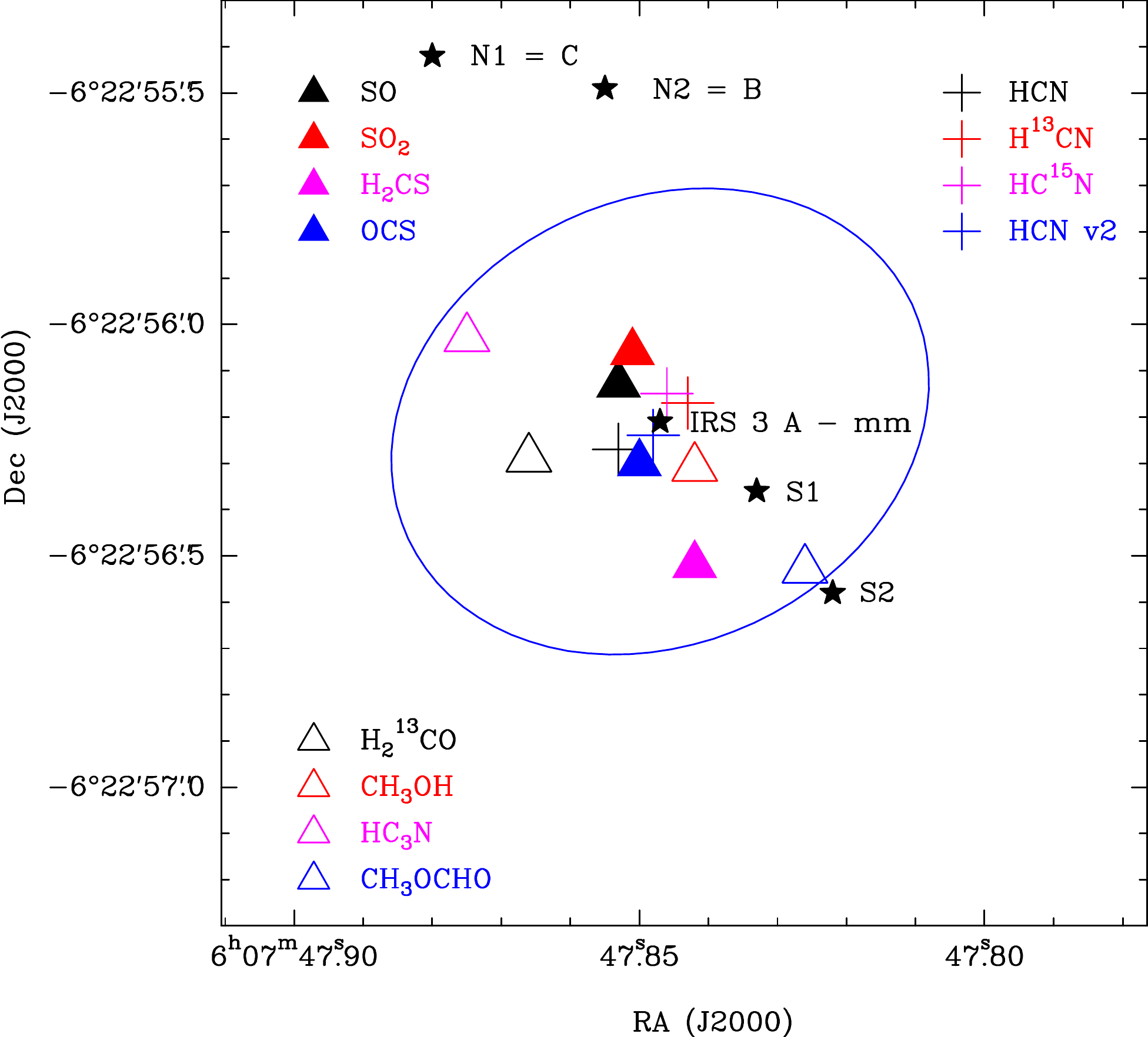}
    \caption{Scheme showing the positions of the emission centroids of the transitions shown in Fig.~\ref{fig:maps}. Blue ellipse indicates a beam of  1.20$"$$\times$0.96$"$ P.A. $-$65$^\circ$. Positions are labeled as in Fig.~\ref{fig:positions}}
    \label{fig:peaks}
\end{figure}

%====================
\section{Observations} 
%====================

\label{sec:obs}
The  IRS~3 massive protostar  was  observed  with  ALMA  (Atacama  Large Millimeter/submillimeter Array) during Cycle 3. The observations were performed on April 2015 within the project 2015.1.00453.S. The project was observed with  the 12m array, with the C40-4 configuration using  36 antennas.
We used the Band 6 receiver in two different tunings corresponding to two different science goals: setup 1 centered at 259.0 GHz and 272.0  GHz in the lower and upper side band, respectively, and setup 2 centered at 249.0 and 262.0 GHz.  A total of 18 spectral windows were observed, each with a spectral resolution $<$0.3 km s$^{-1}$ which allows to fully resolve the profiles of the detected lines (see Table~\ref{table:observations}). Total time on-source is 44.88 min for setup 1 and 53.3 min for setup 2. The phase calibrator was J0607-0834 and the quasar J0522-3627 was observed for bandpass  and flux  calibration. 
The data were calibrated using the ALMA pipeline in the version 4.7 of The Common Astronomy Software Applications\footnote{More information about CASA in http://casa.nrao.edu} (CASA; \citealp{Mullin2007}). Spectroscopic images of all spectral windows were produced in CASA using natural weighting.  In the case of the higher signal-to-noise ratio (SNR) continuum images, we used Briggs weighting (robust=0.5) to achieve higher spatial  resolution. Table~\ref{table:observations} gives the frequency ranges of each spectral window and the final beams.
For analysis purposes we used the  CLASS/GILDAS package \citep{Pety2005}. 
The intensities are given in units of temperature.

%====================
\section{Results} 
%====================
\subsection{Continuum images} 
\label{sec:cont}

We have constructed continuum images at 1.2mm  and 1.1mm  by averaging the channels free from line emission in each band, 
and then merging the visibilities of the 247~GHz + 257~GHz  and 262~GHz + 272~GHz  bands, respectively.  
Total fluxes are S(1.1mm)= 168$\pm$25 mJy and S(1.2mm)= 135$\pm$6 mJy, which implies a spectral index, $\alpha$=3.7$\pm$3.3, consistent with
 the emission being dominated by dust thermal emission.  Fig.~\ref{fig:continuum} shows the higher spatial resolution continuum image at 1.1mm which barely resolves
 two millimeter clumps, hereafter IRS~3~A and IRS~3~B. The position of the millimeter source IRS~3~A is shifted by $\sim$0.38$"$ NE 
 from the NIR position A as observed in K band by \citet{Preibisch2002} (see  Fig.~\ref{fig:positions}). The millimeter source IRS~3~B includes the NIR sources B and C which are indistinguishable with the angular resolution of our observations. We do not detect any millimeter compact emission towards sources  D and E. 
Based on our millimeter observations, we estimate the total gas and dust mass in Mon~R2~IRS~3 using a gray-body model,
\begin{equation}
 M = \frac{ F_{1.1mm} D^2} {B_{1.1mm} (T_{\rm dust})  k_{1.1mm}} 
 \end{equation} 
where M is the gas+dust mass, $D$ is the distance, and we adopt a grain emissivity, $k_{1.1mm}$ = 0.0078 cm$^{2}$ g$^{-1}$, 
which is the value calculated by \citet{ossenkopf1994} for bare grains and dense gas, where gas-to-dust ratio of 100 is assumed.
Adopting  T$_{\rm dust}$ = 200~K  (see Sect.~\ref{sec:mol}), we derive a total mass of $\sim$0.14~M$_\odot$.  
The angular resolution of the 1.1mm image is not enough to cleanly discern between  IRS~3~B and  IRS~3~A.
We have fitted two 2D Gaussians to the 1.1mm image to estimate the contribution of each source to the total flux (see Table~\ref{table:gaussians}). 
We obtain that $\sim$80\% of the total flux comes from IRS~3~A, while  only $\sim$20 \% comes from the northern mm source. However our
fit cannot account for the total flux, proving the existence of a more  complex structure and an extended component.  We adopt the flux and  the half power full size 
of the brightest Gaussian 
to estimate the averaged H$_2$ column  density towards IRS~3~A, N(H$_2$)$\sim$2.6$\times$10$^{23}$ cm$^{-2}$. An independent estimation of N(H$_2$) can be
done from spectroscopic observations. \citet{Dungee2018} estimated N($^{13}$CO)=(1.1$\pm$0.2)$\times$10$^{17}$ cm$^{-2}$ based on the absorption 
NIR lines  towards  IRS~3~A.  Absorption lines are only probing the gas between the continuum source and the observer (in front of the continuum source). 
However, dust continuum emission is tracing the gas along the whole line of sight, i.e.  in front the continuum source and beyond it. Assuming a standard $^{13}$CO abundance of $\sim$2$\times$10$^{-6}$ and that the amount of gas beyond the protostar is the same as that measured by the absorption lines, we derive  N(H$_2$)$\sim$1.1$\times$10$^{23}$ cm$^{-2}$ along the line of sight which is in reasonable agreement with our estimate based on the 1.1mm continuum.

\begin{table*}
	\caption{List of the lines identified in IRS~3}
	\label{table:lines} 
	\centering                 
\begin{tabular}{llrl}   
	\hline \hline    
\textbf{Transition} & \textbf{Rest freq.} & \textbf{E$_{u}$} & \textbf{Log$_{10}$ (A$_{ij}$)} \\
                    & \textbf{(MHz)}      & \textbf{(K)}     & \textbf{(s$^{-1}$)}        \\
\hline\noalign{\smallskip} 
C$_2$H N=3$\rightarrow$2 J=7/2$\rightarrow$5/2 & 262004.26 & 25 & -4.28  \\
\multicolumn{4}{c}{\it Cyanopolynes} \\
HCN 3$\rightarrow$2                  & 265886.43           & 25       & -3.07  \\
HCN $\nu_2$=1 3$\rightarrow$2   & 265852.76           & 1050   & -2.64   \\
H$^{13}$CN 3$\rightarrow$2      & 259011.80           & 25       & -3.11   \\
HC$^{15}$N 3$\rightarrow$2      & 258157.00           & 25       & -3.12    \\ 
HNC 3$\rightarrow$2                  & 271981.14           & 26       & -3.03    \\ 
HC$_3$N 30$\rightarrow$29      & 272884.75           & 203      & -2.79   \\
%CH$_3$CN v$_8$=1     14(0,2)$\rightarrow$13(0, 2)   &  258276.09   & 618.15     &  -2.85   &       \\
\multicolumn{4}{c}{\it Sulphuretted species} \\
SO 7$_6$$\rightarrow$6$_5$             & 261843.68          & 47        & -3.63     \\
SO 6$_6$$\rightarrow$5$_5$             & 258255.83          & 56        & -3.67     \\
SO$_2$ 5(3,3)$\rightarrow$5(2,4)       & 256246.95         & 36        & -3.97    \\
SO$_2$ 10(5,5)$\rightarrow$11(4,8)     & 248830.82        & 112     & -4.66  \\
SO$_2$  31( 9,23)$\rightarrow$32( 8,24)                  & 247169.77       &  654   & -4.50 \\
$^{34}$SO$_2$ 3(3,1)$\rightarrow$3(2,2)  & 247127.39       & 27        & -4.22   \\
OCS                  21$\rightarrow$20      &  255374.46       & 135       &   -4.31   \\
OC$^{34}$S     23$\rightarrow$22       &   272849.96      & 157       & -4.23     \\
H$_2$CS   8(2,7)$\rightarrow$7(2,6)         &  274703.35          &  112      &   -3.54   \\
H$_2$CS   8(3,6)$\rightarrow$7(3,5)  &   274732.97   &     178   &    -3.58   \\
H$_2$CS   8(3,5)$\rightarrow$7(3,4)  &   274734.40   &     178   &   -3.58    \\
\multicolumn{4}{c}{\it CO-H species} \\
%H$_2$CO             10(2,8)$\rightarrow$11(0,11)        & 258296.07      &    241    &  -6.30    \\ 
H$_2$$^{13}$CO 4(1,4)$\rightarrow$3(1,3)              & 274762.11       &    45      & -3.26     \\
HDCO                  4(2,2)$\rightarrow$3(2,1)               & 259034.91      &    63      & -3.44     \\           
CH$_{3}$OH  16(- 2,15)$\rightarrow$15(- 3,13) E     &  247161.95     &  338      & -4.59     \\
CH$_{3}$OH     4( 2, 2)$\rightarrow$5( 1, 5) A           &  247228.58     &   61      &  -4.67 \\
CH$_{3}$OH   16(3,13)$\rightarrow$16(2,14) A         & 248885.48      &  365     & -4.08     \\
CH$_{3}$OH  17( 3,15)$\rightarrow$17( 2,16) A        & 256228.71      &   404    & -4.05     \\   
CH$_{3}$OH   25(- 1,24)$\rightarrow$25(- 0,25) E     &  271933.60    &  775     & -4.26  \\
\multicolumn{4}{c}{\it COMS} \\
CH$_{3}$OCHO    $\nu$ = 1   20( 8,13)$\rightarrow$19( 8,12) E  &  246184.18    &    353   &   -3.72 \\
CH$_{3}$OCHO    $\nu$ = 1   21( 2,19)$\rightarrow$20( 2,18) A  &  246187.02    &    327   &   -3.66 \\
CH$_{3}$OCHO    $\nu$ = 1   20( 7,13)$\rightarrow$19( 7,12) A    &  246233.57    &    344   &   -3.70 \\  
CH$_{3}$OCHO    $\nu$ = 1   20( 7,13)$\rightarrow$19( 7,12) E    &  246274.89    &    343   &   -3.70 \\
CH$_{3}$OCHO      20(11, 9)$\rightarrow$19(11, 8) E                      &  246285.40    &    204   &   -3.80 \\   
CH$_{3}$OCHO      20(11,10)$\rightarrow$19(11, 9) A                     &  246295.13    &    204   &   -3.80 \\
CH$_{3}$OCHO      20(11,9)$\rightarrow$19(11, 8) A                       &  246295.13    &    204   &  -3.80 \\
CH$_{3}$OCHO      20(11,10)$\rightarrow$19(11, 9) E                     &  246308.27    &   204    &  -3.80 \\
CH$_{3}$OCHO    $\nu$ = 1    20( 7,14)$\rightarrow$19( 7,13) E   &  247147.12    &   343    &  -3.69 \\ 
CH$_{3}$OCHO       21(14, 8)$\rightarrow$20(14, 7) E     &  258142.09    &    266   &  -3.84   \\     
CH$_{3}$OCHO       21(13, 8)$\rightarrow$20(13, 7) E     &  258274.95    &    248   &   -3.79  \\           
CH$_{3}$OCHO       21(13, 8)$\rightarrow$20(13, 7) A     &  258277.43    &    248   &   -3.79  \\        
CH$_{3}$OCHO       21(13, 9)$\rightarrow$20(13, 8) A     &  258277.43    &    248   &   -3.79 \\        
CH$_{3}$OCHO       21(13, 9)$\rightarrow$20(13, 8) E     & 258296.30     &    248   &   -3.79  \\        
CH$_{3}$OCHO   $\nu$ = 1      21(7,14)$\rightarrow$20(7,13) A    & 259003.87     &    356   &   -3.63 \\
CH$_{3}$OCHO   $\nu$ = 1      21(7,14)$\rightarrow$20(7,13) E    & 259025.83     &    356   &   -3.63 \\
CH$_{3}$OCHO      22( 8,15)$\rightarrow$21( 8,14) A     & 273135.04     &    192   &    -3.57 \\        
CH$_{3}$OCHO      22( 8,14)$\rightarrow$21( 8,13) E    & 273142.74     &    192   &    -3.58 \\      
CH$_{3}$OCHO      22( 8,15)$\rightarrow$21( 8,14) E    & 273151.24    &    192   &    -3.58 \\        
CH$_{3}$OCHO      22( 8,14)$\rightarrow$21( 8,13) A    & 273180.43    &    192   &    -3.57 \\   
CH$_{3}$OCHO    $\nu$ = 1   22(6,17)$\rightarrow$21(6,16) E  & 273653.25     &    361   &    -3.55 \\   
CH$_{3}$OCH$_3$  16( 1,15)-15( 2,14) EE   &  273107.18  &   127   &   -4.06   \\
CH$_{3}$OCH$_3$  15( 5,10)-15( 4,11) EE   &  261248.11   &   144  &    -4.07  \\
CH$_{3}$OCH$_3$  15( 5,10)-15( 4,11) AA   &  261250.49   &   144  &    -4.06  \\
CH$_{3}$OCH$_3$  15( 5,11)-15( 4,12) EE   &  261956.62   &   144  &   -4.06 \\
\hline\noalign{\smallskip}
\end{tabular}

\end{table*}

\subsection{Spectroscopic observations}
\label{sec:lineiden}
To study the chemical content of IRS~3, we have extracted the interferometric spectra
towards IRS~3~A and IRS~3~B. Eye  inspection reveals that the spectra towards IRS~3~A are more crowded than those towards IRS~3~B (see Fig.~\ref{fig:continuum}),
thus confirming that the hot core is associated with the mm source  IRS~3~A.  
The lines have been identified using the $"$The Cologne Database for 
Molecular Spectroscopy$"$ (CDMS) 	\citep{Muller2005, Endres2016} and the "Jet Propulsion Laboratory Line catalogue" (JPL) \citep{Pickett1998}. 
We have adopted the 4$\sigma$ level in integrated intensity as the detection limit. 

The list of identified lines is shown in Table~\ref{table:lines}. 
We can divide the detected species in different families according with their chemical similarities: 
(i) the small hydrocarbon C$_2$H ; (ii) the nitrogenated  carbon chains
HCN, HNC, H$^{13}$CN, HC$^{15}$N, and HC$_3$N; (iii) the  S-bearing species SO,  SO$_2$,  $^{34}$SO$_2$, OCS,  H$_2$CS,
and OC$^{34}$S; and  (iv) the organics CH$_3$OH, HDCO, and H$_2$$^{13}$CO, and  COMs such  as methyl formate (CH$_3$OCOH)
and dimethyl ether (CH$_3$OCH$_3$). The detections of these species are robust. 
We have checked the identification of the S-bearing species  by searching for the $^{34}$S  isotopologues to test that
the (non-)detections of all the $^{34}$S isotopologues are compatible with a  standard ratio of $^{32}$S/$^{34}$S=22.5 \citep{Wilson1994, Chin1996}. 
The most doubtful identification is dimethyl ether. We have identified 4 lines that can be assigned to CH$_3$OCH$_3$ but three of them overlapped.  
In order to confirm (reject) this  detection,
we produced a synthetic spectrum assuming Local Thermodynamic Equilibrium and T$_{ex}$=170~K and
checked that the obtained spectrum is compatible with the observed spectra (see  Sect.~\ref{sec:mol} and Figs.~\ref{fig:espectros1} to Fig.~\ref{fig:espectros2}). 
The COMs CH$_3$OCHO and  CH$_3$OCH$_3$ are only  
detected towards IRS~3~A  proving  that the hot core is associated to this protostar.

Different line profiles can be observed in the spectra shown in Figs.~\ref{fig:espectros1} to ~\ref{fig:espectros2}, revealing that the 
observed lines come from different regions along  the line of sight. 
The C$_2$H 3$\rightarrow$2 and HNC 3$\rightarrow$2 lines present intense absorptions towards the continuum compact source. A small feature in absorption is also detected at the frequency of the c-C$_3$H$_2$ 4$_{4,1}$$\rightarrow$3$_{3,0}$ line (see Fig.~\ref{fig:espectros2}). 
Previous works in the (sub-)millimeter range showed that HNC, C$_2$H and c-C$_3$H$_2$ are
abundant species in the surrounding molecular cloud with bright emission of the HNC 3$\rightarrow$2 and C$_2$H 3$\rightarrow$2 lines \citep{Pilleri2013,Trevino2014,Trevino2019}. 
 These absorptions are more likely produced by the cold gas in the envelope which absorbs the continuum and line emission from the hot core, and are not of interest for the goals of
 this paper. It should be noticed that our observations are filtering the extended emission, thus producing
negative contours in the emission of these molecules also far from the compact hot core. Signatures of missed flux  are also found in 
the HCN 3$\rightarrow$2,  H$^{13}$CN 3$\rightarrow$2, and to a lesser extent in the SO 6$_6$$\rightarrow$5$_5$  and 7$_6$$\rightarrow$6$_5$ images.
Contrary to C$_2$H and HNC, these molecules present intense compact emission towards IRS~3~A and we keep them in our study.  We recall that we are interested in the
emission coming from the hot core that is not affected by the spatial filtering.
Linewidths of $\sim$4~km~s$^{-1}$ to $\sim$10~km~s$^{-1}$ are measured in the detected lines. These linewidths are not correlated with
the line excitation conditions. In fact, the more abundant species such as HCN, SO,  and  CH$_3$OH  present wider profiles suggesting that they are
the consequence of higher opacities and hence, higher sensitivity to the lower gas column densities measured at high velocities.  A different case would be
the vibrational excited  line of HCN that is  indeed probing the hottest  regions of  this disk and outflow system.
In addition to larger linewidths, the HCN 3$\rightarrow$2,   SO 6$_6$$\rightarrow$5$_5$  and 7$_6$$\rightarrow$6$_5$ lines show high 
velocity wings which extend from $-$15~km~s$^{-1}$ to $\sim$22~km~s$^{-1}$,  suggesting that these lines could be partly tracing the outflow detected in IRS~3 by \citet{Dierickx2015}.
It is remarkable the non-detection of HC$^{18}$O$^+$  and N$^{15}$NH$^+$ J=3$\rightarrow$2 lines. These molecular ions are not expected to be abundant in 
hot cores (see e.g. \citealp{Crockett2014}).
The high gas temperatures measured in this region \citep{vandertak2003, Dierickx2015, Dungee2018}, and the detection of methyl formate confirm the existence of a hot core associated with Mon~R~2 IRS~3, already suggested by  \citet{Dierickx2015} that now we can identify with IRS~3~A.

\begin{figure*}
    \centering
    \includegraphics[width=0.8\textwidth]{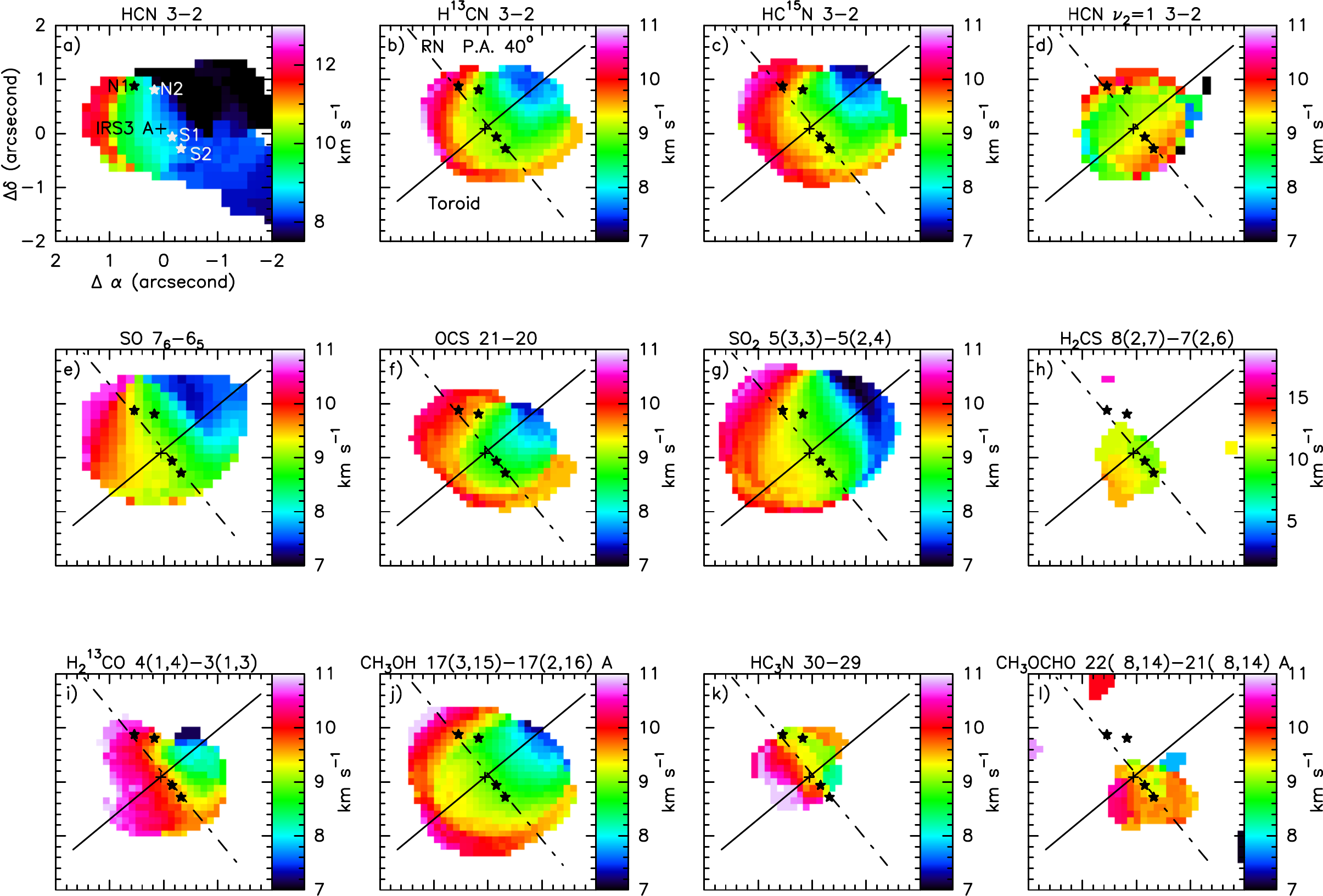}
    \caption{First-moment maps of the most intense lines detected in ALMA data. Straight lines in panel b) indicate the directions  of the Position-Velocity diagrams in the directions perpendicular to the NIR nebula (toroid) and along it (RN, P.A. 40$^\circ$). The beams are the same as in Fig.~\ref{fig:maps}. }
    \label{mapas-mom1}
\end{figure*}

\subsection{Moment maps}
\label{sec:mom}

The integrated intensity maps (zeroth-moment maps) of the most intense lines are shown in Fig.~\ref{fig:maps}. 
We exclude C$_2$H 3$\rightarrow$2 and HNC 3$\rightarrow$2 in these maps because, as commented before, their absorptions
are coming from the surrounding envelope. Since this  paper is focused on the kinematics and chemistry of the hot core, we are not 
going to further discuss these species. The emission of the rest of lines is centered on the vicinity of IRS~3~A, and is, in general, more extended
than the dust emission. There are some differences between the spatial distributions of the different species.  
It is remarkable that the emission of  H$_2$CS and CH$_3$OCHO are more intense 
in the southern part of the NIR nebula, with the emission of the CH$_3$OCHO line being maximum towards S2. In contrast, the emission of
HC$_3$N is more intense in the northern nebula. Fig.~\ref{fig:peaks} shows a scheme with the locations of the emission peaks  of the
different  molecules. Most of the observed species  peak in a region R$<$0.3" from IRS~3~A. Interestingly the emission  of HCN and 
their isotopologues show the best agreement with the position of the continuum peak (labeled "IRS~3~A-mm" in Fig.~\ref{fig:peaks}).  
The peak of the CH$_3$OCHO is located $\sim$0.5$"$ SW to IRS~3~A-mm, that is about a half of the HPBW of our observations.  
The peak of  the HC$_3$N emission is displaced by  $\sim$0.5$"$ NE from  IRS~3~A-mm. 
These displacements prove the chemical complexity of the IRS~3 hot core with at least two components, the brightest one associated to the torus that is forming the reflection nebula, and the second one originated in the NIR reflection nebula. 

Relevant information on the structure and kinematics of the molecular gas  can  also be obtained from the first-moment maps shown in 
Fig.~\ref{mapas-mom1}. First-moment maps show the  intensity-weighted average velocity at every position, hence providing information on the
emitting gas kinematics. Most of the first-moment maps present a clear SE-NW velocity gradient in the v$_{lsr}$=7$-$11~km~s$^{-1}$ velocity 
range. The protostar IRS~3~A cannot be surrounded  by a thick spherical envelope because it is associated to a bipolar NIR nebula.
The detected velocity gradient is  consistent with the emission coming from a toroidal envelope and/or a disk that is  rotating in the direction perpendicular
 to the axis defined by the NIR nebula
(the direction of the reflection nebula associated to IRS~3~A is P.A.$\sim$ 40$^\circ$ and we will refer to it as "RN", hereafter). 
The v$_{lsr}$=7$-$11~km~s$^{-1}$ velocity gradient is not detected in the HCN $\nu_2$=1 J=3$\rightarrow$2 and CH$_3$OCHO first-moment maps.
We recall that the CH$_3$OCHO emission peaks towards the southern reflection nebula, suggesting a different origin. 
Moreover, the emission is red-shifted relative to the systemic velocity, v$_{sys}$ $\sim$ 9 km s$^{-1}$, consistent with the
interpretation that this line is  associated to the walls of the cavity excavated by the CO outflow reported by  \citet{Dierickx2015}). 
In the case of  HCN $\nu_2$=1, although it is very marginal with the angular resolution of the observations, 
there seems  to be a  velocity gradient along the reflection nebula.

\begin{figure*}
    \centering
    \includegraphics[width=0.9\textwidth]{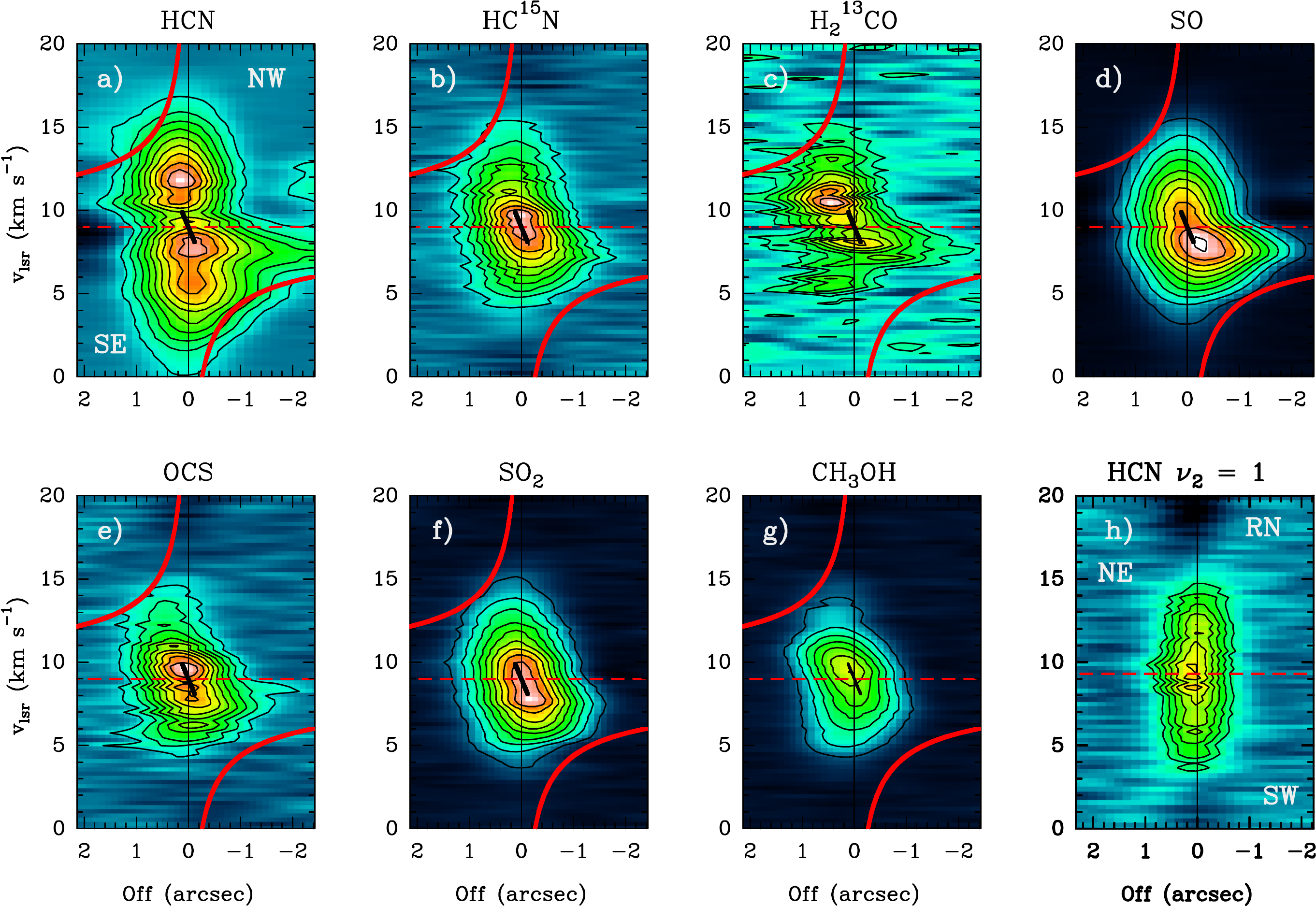}
    \caption{
    a), b), c), d), e), f), and g): Position-Velocity (PV) diagrams along the axis perpendicular to the nebula ({\it toroid} in  Fig.~\ref{mapas-mom1}). The red solid lines indicate the Keplerian velocity  for a mass of 40~M$_\odot$ and an inclination i=45$^{\circ}$ while  the solid black line show  the velocity pattern for a solid-body model. h) PV diagram along the direction of  the NIR nebula ({\it RN} in Fig.~\ref{mapas-mom1}) of the HCN $\nu_2$=1 3$\rightarrow$2 line.  First contour and  step  (both share the same value) in each panel are: a) 3.72~K; b) 0.74~K; c) 0.25~K; d) 5.00~K;  e) 0.56~K; f) 1.76~K; g) 1.77~K;  h) 1.00~K. }
    \label{fig:cortes}
\end{figure*}

 To further explore the kinematics of the molecular gas and distinguish between a toroidal envelope from the circumstellar disk,  
 we have performed Position-Velocity (PV) diagrams in the direction perpendicular 
 to the reflection nebula ("toroid" in Figs.~\ref{mapas-mom1}b).
In Fig.~\ref{fig:cortes}, we compare these PV-diagrams with the velocity patterns expected in the case of Keplerian rotation (red line)
and solid-body rotation (black line). Only  HCN and H$_2^{13}$CO present  a "butterfly" shape that could be interpreted as Keplerian 
rotation. For comparison we have plotted the Keplerian curve expected for M=40 M$_\odot$ and $i$ = 45$^\circ$. 
\citet{Preibisch2002} provided an estimation of the masses of the NIR protostars. Assuming that  the NIR  sources  A, B and C are driving the 
 toroid rotation, the stellar mass would be 25-37 M$_\odot$. Yet, we can be missing an embedded protostar(s) that 
 remains invisible at NIR  wavelengths.  We adopted 40~M$_\odot$  as a first approximation to the total stellar mass. The Keplerian 
 velocity  depends  on the $M_{stars} \cdot sin(i)$ where $i$ is the unknown inclination angle of the toroid. Therefore, there  is a  degeneration 
 between the  total stellar mass and the  inclination angle.   We find a reasonable agreement  with HCN and H$_2^{13}$CO observations 
 assuming $M_{stars}=40 M_\odot$ and  $i=45^\circ$. It should be noticed that
a  small change in the stellar mass could be compensated by a  slight change the inclination angle. The match between the Keplerian curves 
thus obtained and the PV diagrams of  HCN and H$_2^{13}$CO is reasonable in the NW region 
but poor in the SE. As commented above, we cannot discard some contribution of an unresolved component of the molecular outflow 
detected by \citet{Dierickx2015} to the high velocity gas. 
The PV-diagram of the rarer isotopologue HC$^{15}$N is more consistent with solid-body rotation. Solid-body rotation is expected if the
emission is coming from a rotating ring.  In the cases of unresolved disks, where the sensitivity is not enough to detect the highest rotation velocities
expected close to the star, a Keplerian disk and solid-body rotation would be difficult to differentiate.
Thus, the non detection of  HC$^{15}$N
at high velocities might be a sensitivity problem since HCN/HC$^{15}$N$\sim$ 300 \citep{Hily2013}.  
Solid-body rotation is also compatible with the PV diagrams of SO, SO$_2$, OCS, CH$_3$OH, and HC$_3$N.  
One would expect chemical gradients within the ring with the
N(SO$_2$)/N(SO) and N(CH$_3$OH)/N(H$_2$CO) ratios increasing inwards \citep{Gieser2019} but the limited  angular resolution and 
sensitivity of our observations do not allow to discern it.

The emission of the  HCN $\nu_2$=1 J=3$\rightarrow$2 line is centered on IRS~3~A with linewidths of $\sim$10~km~s$^{-1}$.
The HCN $\nu_2$=1 emission is unresolved in the direction of the toroid, but presents some extension to the  north along the outflow direction
(see Fig.~\ref{mapas-mom1}).
In order to explore the details of the kinematics of the hot gas traced by the vibrationally excited HCN line, we performed a PV diagram along the outflow
direction  ("RN" in Figs.~\ref{mapas-mom1}b). The PV diagram shows an essentially  unresolved emission without any clear evidence  of a velocity
gradient along this axis. On the other hand, the high excitation conditions of the HCN $\nu_2$=1 line and the large linewidths observed would also be consistent with the emission
 coming from the inner disk region. Given that the emission of the   HCN $\nu_2$=1 is  not resolved,  we could not disentangle between these two physical processes.

\section{Chemical content}
\label{sec:mol}

The estimation of accurate molecular column densities requires the knowledge of the excitation conditions. The excitation temperature
can be easily derived using the rotation diagram technique as long as several transitions are observed. We have identified several lines 
of SO, SO$_2$, CH$_3$OH, and CH$_3$OCHO in 
our interferometric spectra (see Table~\ref{table:lines}). The CH$_3$OCHO lines are weak and there is large overlapping between different transitions, 
hindering the estimation of the rotation temperature.  The two SO lines have similar $E_u$ precluding an accurate estimate of T$_{rot}$.
Therefore, we have used the SO$_2$ and CH$_3$OH lines to estimate the gas temperature. We obtained  T$_{rot}$ = 175$_{-16}^{+15}$ K and N(SO$_2$)=3.4$_{-0.5}^{+0.6}$ $\times$ 10$^{16}$ cm$^{-2}$ using the three SO$_2$ lines observed towards IRS~3~A (see Table~\ref{table:cd}). A slightly  lower value of the rotation temperature, T$_{rot}$ = 129$_{-9}^{+10}$ K, and  N(CH$_3$OH)=1.1$_{-0.2}^{+0.3}$ $\times$ 10$^{17}$ cm$^{-2}$,  are found using the methanol lines. These temperatures are consistent with those previously obtained by other authors.
\citet{Boonmanvd2003} derived T$_{ex}$=250$_{-100}^{+200}$K based on  the $\nu_2$ band of  H$_2$O.  Excitation temperatures 
of 110$\pm$20~K and 125$\pm$35~K for CH$_3$OH and SO$_2$, respectively,  were estimated by  \citet{vandertak2003} on the basis of single-dish observations.
\citet{Dierickx2015} estimated T$_{ex}$= 126$\pm$ 22 K toward IRS~3 from the CH$_3$CN line emission. 
\citet{Dungee2018} obtained T$_{ex}$=234$\pm$15~K from $\nu_3$ ro-vibrational band of SO$_2$.
We can conclude that the observed excitation temperatures ranges from $\sim$110 K using millimeter single-dish observations to T$_{ex}$=250~K when using the
infrared ro-vibrational bands, suggesting a temperature gradient along the line of sight.

The SO$_2$ column density derived from our data is in good agreement with that obtained by \citet{Dungee2018}. However, we do not obtain a good fit to the three millimeter lines observed,  which is the consequence of the steep gas temperature gradient along the line of sight  (see Figs.~\ref{fig:espectros1}). We cannot obtain a good fit to all the methanol lines using a single rotation temperature, either. Given the small number of transitions observed, we do not consider that
we could obtain realistic results using a more complex fitting with several rotation temperatures.
The molecular column densities of the rest of species have been derived by fitting the lines observed towards IRS~3~A assuming Local Thermodynamic Equilibrium (LTE), which is a reasonable assumption for the high densities (n(H$_2$)$>$10$^7$~cm$^{-3}$) expected in a hot core. Taking into account the high SNR ($>>$10) in most of the lines, the main uncertainties in the derived column densities would come from the adopted rotation temperatures. We estimate these uncertainties by calculating the column densities with
two temperatures, 129~K -170~K, that are representative
of the temperatures derived using different molecular tracers (see Table~\ref{table:cd}). The uncertainties thus obtained are of $\sim$20\%-25\% for most species, except
for the vibrationally excited HCN for which the differnce is of more than one order of magnitude due its high excitation conditions. In Figs.~\ref{fig:espectros1} and ~\ref{fig:espectros2}, 
we show the results for T$_k$=170~K. The quality of the fit is essentially the same when using T$_k$=129~K and the column densities derived with this assumption.
We have also derived an upper limit to the column densities of HC$^{18}$O$^+$ and N$^{15}$N$^+$ column densities based on the non-detection of the HC$^{18}$O$^+$ J=3$\rightarrow$2 lines. Our limit indicates that N(HCN)/N(HCO$^+$)$\sim$8. This value has been calculated from thee H$^{13}$CN and HC$^{18}$O$^+$
 column densities assuming  $^{12}$C/$^{13}$C=60 \citep{Savage2002} and $^{16}$O/$^{18}$O=550 \citep{Wilson1994}.  This high N(HCN)/N(HCO$^+$) ratio is found in the innermost regions of the Class I disks and is interpreted as the result of the high densities and temperatures prevailing in these regions \citep{Agundez2008, Fuente2012, Crockett2014, Fuente2020}.  

Molecular abundances have been estimated  adopting the value derived from our millimeter observations, N(H$_2$)=2.6$\times$10$^{23}$ cm$^{-2}$ (see Table~\ref{table:cd}).
Abundances of $\sim$ a few 10$^{-7}$ are measured for HCN, CH$_3$OH, and SO$_2$. Our estimated SO$_2$ abundance is 5 times lower than that derived by \citet{Dungee2018} mainly due to the higher adopted value of N(H$_2$). We recall that the CH$_3$OH lines are very likely optically thick and the derived abundance a lower limit to the real value.
Following our calculations, SO$_2$ accounts for a significant fraction, $\sim$1\%, of the sulphur budget assuming the solar abundance, S/H=1.5$\times$10$^{-5}$ \citep{Asplund2005}. The abundance of SO$_2$ is a factor $>$5 higher than those of SO, OCS, and H$_2$CS. It would be noticed that our calculations of the column densities of SO$_2$ and OCS are not heavily affected by opacity effects. We have fitted  the lines of the isotopologues $^{34}$SO$_2$, OC$^{34}$S
and derived N(XS)/N(X$^{34}$S) ratios $\sim$22.5 (see Table~\ref{table:cd}, and Fig.~\ref{fig:espectros1} and ~\ref{fig:espectros2}), proving that the lines of the
main isotopologues are optically thin. 
Unfortunately,  we do not have intense $^{34}$SO lines in our interferometric spectra to estimate the opacity of the SO lines.
We do not have interferometric observations of the abundant species, H$_2$S and CS, that would be useful to complete the inventory of S-bearing molecules.
The sulphur chemistry in Mon R2  IRS~3 was previously investigated by
\citet{vandertak2003} using single-dish observations of a wider set of S-bearing molecules including H$_2$S. They found that SO$_2$ was the most abundance 
S-bearing molecule, suggesting that SO$_2$ is the most abundant "observable" S-bearing molecule in this protostar and the main sulphur reservoir in gas phasee. 
%Assuming that SO$_2$ is the main molecular reservoir of sulphur, we would conclude that only  $\sim$1\% of sulphur is in gas phase
%in the hot core IRS~3~A.

\section{Discussion}
\label{sec:dis}

High spatial resolution observations show that hot cores are complex objects in which molecular species present differentiated 
spatial distributions \citep{Zerni2012, Qin2015, Rivilla2017, Bonfad2017, Palau2017, Allen2017, Tercero2018, Mininni2020,Gieser2021}. 
One well-known example is the hot corino IRAS 16293-2422 which is formed by two protostars separated 
by 5$"$  ($\sim$14~au), and usually referred to as A and B, 
IRAS 16293–2422 A being the most intense in the emission of S-bearing species  \citep{Chandler2005, Caux2011}. Recent
observations at an angular resolution of 0.1$"$ ($\sim$700~au) revealed that IRAS 16293–2422 A is itself a multiple source with complex kinematics 
and chemistry \citep{Oya2020}. Interferometric studies using large millimeter telescopes like ALMA are therefore required to have a more precise view of the molecular
chemistry in these objects. We present high spatial resolution ALMA observations of the hot core  Mon~R~2~IRS~3.
Our interferometric images show that the Mon~R~2~IRS~3  is formed by two mm cores, IRS~3~A and IRS~3~B. 
The molecular emission is mainly associated with IRS~3~A, which seems to be the one that harbors the hot core.
The protostar  IRS~3~A itself present a complex  structure in which we identify at least two components: 
i) most species such as HCN (and its isotopologues), SO, SO$_2$, OCS, H$_2^{13}$CO and CH$_3$OH comes from the rotating ring that is forming the
nebula; ii) the emission of the H$_2$CS and CH$_3$OCHO lines come from the southern NIR nebula; iii) the emission of HC$_3$N is shifted to the north. 
In addition, the detailed analysis of the PV diagrams suggests that there  could be
chemical differentiation within the ring, with N(SO)/N(SO$_2$) and N(CH$_3$OH)/N(H$_2$CO) increasing inwards. However, we should put a word of caution in this result
since the SO and CH$_3$OH lines might be optically thick.

Chemical models predict that the abundances of sulphur-bearing species are expected to be enhanced 
in hot cores (e.g., \citealp{Charnley1997, Hatchell1998, Viti2004, Herpin2009, Wakelam2011, Vidal2018}). 
During the collapse, the bulk of the sulphur reservoir is locked onto grain mantles (see e.g. \citealp{Vidal2017, Navarro2020}).
These species are released to the gas phase again when 
dust temperatures increase to $>$100 K and the ice mantles are evaporated in the hot core/corino stage.
Then, an active S-chemistry is initiated where SO$_2$ is a direct product of SO, through the radiative 
association of O and SO and the neutral-neutral reaction of SO with OH, making the SO/SO$_2$ ratio decrease as  the
hot core/corino evolves. \citet{Dungee2018} observed the $\nu_3$ ro-vibrational band of SO$_2$ confirming the existence of hot molecular gas (T$_k$=234$\pm$15 K) 
towards IRS~3, with a high SO$_2$ abundance ($\sim$ a few 10$^{-7}$). 
The detection of SO, OCS, H$_2$CS and SO$_2$ in our interferometric images, provides a more complete view of the sulphur chemistry in this hot core.
The N(SO)/N(SO$_2$) ratio is $\sim$0.29 in IRS~3~A similar to that observed in the Orion hot core and AFGL~2591~VLA~3
(see Table~\ref{table:comp}). Similar values of  N(SO)/N(SO$_2$) are also found in a significant fraction of the massive protostellar objects
imaged within the The NOrthern Extended Millimeter Array (NOEMA) large program “Fragmentation and disk formation during high-mass star
formation" (CORE) \citep{Gieser2021}. Based only in this ratio, one could think that the IRS~3~A presents the characteristic chemistry of an 
evolved hot core such as Orion KL. However we find a great difference when comparing the abundances of COMs in IRS~3~A and Orion, the COMs being
more abundant in Orion. For instance, N(SO$_2$)/N(CH$_3$OCHO)$\sim$0.38 in Orion and N(SO$_2$)/N(CH$_3$OCHO)$\sim$11 in IRS~3~A (see Table~\ref{table:comp} ). 
One possibility is that the CH$_3$OCHO lines are optically thick and we are underestimating their column densities. Another possibility is that 
 the N(SO)/N(SO$_2$) ratio is overestimated  because the SO lines are optically thick, and the hot core is in an earlier evolutionary stage.
A more complete study of  chemistry  of IRS~3~A , including isotopologues,  is required  for an accurate  comparison.

We should also consider the possibility that the high abundance of SO$_2$ in IRS~3~A is produced by the shocks associated with outflow(s) of the region.
Indeed shock chemistry successfully explain the gas-phase SO$_2$ abundance measured toward the
Orion Plateau, which shows very broad lines indicative of shocks generated by Orion IRC 2 outflows \citep{Blake1987,Esplugues2013, Esplugues2014,Crockett2014}. Shocks can sputter dust grains \citep{May2000,Holdship2016}, leading to the release of more
sulphur and subsequent SO$_2$ formation. Recent interferometric observations of S-bearing species towards the outflows associated with  L~1157 \citep{Feng2020}
and NGC 1333 IRAS 4 \citep{Taquet2020} showed important gradients in the SO/SO$_2$ ratio along the outflows that were interpreted in terms of time evolution, 
with SO$_2$ being more abundant at later stages \citep{Taquet2020, Feng2020}.  
Indeed, shocks could also produce enhanced abundances of COMs in intermediate-mass and massive protostars (see, e.g., \citealp{Palau2011, Palau2017}).
We cannot therefore discard that the SO$_2$ emission towards IRS~3~A has some contribution from shocked gas close to the star.
However, the SO$_2$ lines toward Mon~R~2 IRS~3~A are substantially narrow $\Delta V \sim$ 4 km s$^{-1}$,  compared with the high velocities observed
in the bipolar outflow detected by \citet{Dierickx2015}. Moreover, the PV diagrams of the S-bearing species are similar to those
of HC$^{15}$N without any signature of a different origin (see Fig.~\ref{fig:cortes}). Our results are therefore more consistent with the interpretation that sulphur
has been released from the ices due to radiative heating (or perhaps mild shocks)  rather than sputtered by strong shocks. 

Although a large theoretical and observational effort has been undertaken in the last five years to understand sulphur chemistry, there is still a big debate about the main 
sulphur reservoirs in gas phase and volatiles, and eventually about the sulphur elemental abundance in  molecular clouds \citep{Fuente2016, Vidal2017, Fuente2019, Legal2019, Laas2019, Navarro2020, Shing2020, Bulut2021}).
According to chemical models, SO and SO$_2$ are the main gas-phase sulphur reservoirs in gas phase under the conditions of high temperature and density prevailing in hot cores \citep{Esplugues2014, Vidal2017, Gieser2019}. This is consistent with our findings since the abundance of SO$_2$ is a factor $>$5 higher than those of SO, OCS, and H$_2$CS. Summing the abundances of all S-bearing molecules detected towards IRS~3~A, we can account only for $\sim$ 1 \% of the
total sulphur budget. Even assuming that the abundances of H$_2$S and CS are similar to that of SO$_2$, only $\sim$ 3\% of  the sulphur 
is in gas molecules, i.e. , $>$ 95\% should be  locked in ices or refractories.  However, one would not expect icy water at the high temperatures measured in this source. 
One  possibility is that sulphur remains locked in allotropes (S$_2$, S$_3$, ..., or the most  stable S$_8$) as suggested by \citet{Jimenez-Escobar2011}.  
\citet{Shing2020} have shown that the inclusion of cosmic ray-driven radiation chemistry and fast non-diffusive bulk reactions in grain surface chemistry
lead to a reduction in the abundance of solid. H$_2$S and HS, and a significant increase in the abundances of OCS, SO$_2$, and allotropes such as S$_8$ in the ice. 
Experiments and theoretical  work is still needed to determine the solid reservoir of sulphur.

\begin{table}
	\caption{Molecular column densities and abundances}
	\label{table:cd} 
%	\centering                 
\begin{tabular}{lllcl}   
	\hline \hline    
	\multicolumn{1}{c}{Mol} & \multicolumn{2}{c}{Column densities (cm$^{-2}$)}  &   
	 \multicolumn{1}{c}{$\Delta V$(km~s$^{-1}$)}  & \multicolumn{1}{c}{Abundance$^1$} \\ 
	 	\multicolumn{1}{c}{} & \multicolumn{1}{c}{T$_k$=129~K}  &    \multicolumn{1}{c}{T$_k$=170~K} &
	 \multicolumn{1}{c}{}  & \multicolumn{1}{c}{relative  to H$_2$} \\ 
 \hline\noalign{\smallskip} 
HCN                       &    8.7$\times$10$^{14}$       &  1.1$\times$10$^{15}$       &  10.0  &   4.2$\times$10$^{-9}$    \\
HCN $\nu_2$=1     &    3.5$\times$10$^{17}$       &  6.0$\times$10$^{16}$       &    10.0  &   2.3$\times$10$^{-7}$       \\
H$^{13}$CN           &   2.4$\times$10$^{14}$        &   3.0$\times$10$^{14}$       &    7.0  &   1.1$\times$10$^{-9}$       \\  
HC$^{15}$N           &   9.0$\times$10$^{13}$        &   1.1$\times$10$^{14}$       &    6.0  &   4.2$\times$10$^{-10}$    \\ 
HC$_3$N               &    1.7$\times$10$^{13}$       &    1.5$\times$10$^{13}$      &    4.0  &   5.8$\times$10$^{-11}$    \\  
SO                          &    8.0$\times$10$^{15}$       &    1.0$\times$10$^{16}$      &    7.0  &   3.8$\times$10$^{-8}$     \\  
%$^{34}$SO        &    3.5$\times$10$^{14}$   &   3.6$\times$10$^{14}$   &   1.2$\times$10$^{-9}$ \\
SO$_2$$^*$                  &             ...                                 &    3.4$\times$10$^{16}$      &   6.0   &   1.3$\times$10$^{-7}$  \\                                
$^{34}$SO$_2$     &            ...                                  &   1.0$\times$10$^{15}$     &   6.0    &   3.8$\times$10$^{-9}$  \\   
OCS                        &    3.0$\times$10$^{15}$       &     2.5$\times$10$^{15}$     &   6.0    &   9.6$\times$10$^{-9}$  \\ 
OC$^{34}$S            &    8.8$\times$10$^{13}$      &      8.7$\times$10$^{13}$    &   6.0   &   3.3$\times$10$^{-10}$  \\ 
H$_2$CS                &     9.0$\times$10$^{13}$     &     8.0$\times$10$^{13}$    &   4.0   &    3.1$\times$10$^{-10}$  \\                          
H$_2$$^{13}$CO    &   4.5$\times$10$^{14}$       &   6.0$\times$10$^{14}$     &   6.0   &     2.3$\times$10$^{-9}$  \\
HDCO                       &  1.2$\times$10$^{14}$      &  1.5$\times$10$^{14}$     &  4.0     &  5.7$\times$10$^{-10}$  \\
CH$_3$OH$^*$        &   1.4$\times$10$^{17}$    &           ...                          &   6.0     &  5.4$\times$10$^{-7}$   \\
CH$_3$OCHO          &   2.5$\times$10$^{15}$    &   3.0$\times$10$^{15}$    &  4.0   &  1.1$\times$10$^{-8}$  \\
CH$_3$OCH$_3$     &   8.0$\times$10$^{14}$     &  1.1$\times$10$^{15}$     &  4.0   &  4.2$\times$10$^{-9}$ \\
HC$^{18}$O$^+$$^\#$     &    $<$ 1.6$\times$10$^{12}$  &     $<$ 2.0$\times$10$^{12}$ &   4.0  &  $<$8$\times$10$^{-12}$   \\
N$^{15}$NH$^+$$^\#$      &    $<$ 1.5$\times$10$^{12}$  &    $<$ 1.8$\times$10$^{12}$   &   4.0 &  $<$7$\times$10$^{-12}$   \\
\hline\noalign{\smallskip}
\end{tabular}

\noindent
Notes: $^1$ Adopting N(H$_2$)=2.6$\times$10$^{23}$ cm$^{-2}$ derived  in Sect.~\ref{sec:cont} and column densities derived with T$_k$=170~K, except for methanol;  
$^*$ We have derived these numbers using the rotational diagram  technique  (see text  for details);
$^\#$ The rms has been derived in a channel of $\Delta v$=1~km~s$^{-1}$.  Then, the upper limit is calculated assuming  a linewidth of 4 km~s$^{-1}$
and imposing T$_{\rm b}$$<$3$\times$rms.
\end{table}

\begin{table*}
	\caption{Comparison  with other sources}
	\label{table:comp}     
\begin{tabular}{lllllll}   
	\hline \hline    
\multicolumn{1}{c}{Hot core} & \multicolumn{1}{c}{N(SO)/N(SO$_2$)} & \multicolumn{1}{c|}{N(OCS)/N(H$_2$CS)}  &
\multicolumn{1}{c}{N(CH$_3$OCHO)/N(CH$_3$OH)}   &  \multicolumn{1}{c}{N(SO$_2$)/N(CH$_3$OCHO)}  &  \multicolumn{1}{c}{N(H$_2$CS)/N(CH$_3$OCHO)} 
 \\ \hline\noalign{\smallskip} 
IRS 3 A                           &  $\sim$0.29   &    $\sim$31    &   $\sim$0.02   &  $\sim$11         &  $\sim$0.03  &    \\
%IRAS+16293-2422         &  $\sim$3        &                      &                         &                          &                          &                      \\
NGC~7129$-$IRS~2         &  $<$125        &   $\sim$7        &  $\sim$0.014    &   $\sim$0.16    &    $\sim$0.38     \\
Orion                                 &  $\sim$1        &    $\sim$13     &  $\sim$0.26      &   $\sim$0.38    &    $\sim$0.01      \\
AFGL 2591 VLA 3             &   $\sim$0.5     &       ...                 &  $\sim$0.05     &   $\sim$77       &        ...                   \\   
%L1157 B1                      &   $\sim$1        &   $\sim$0.3-2  &                        &                         &                          &                          &                      \\
\hline\noalign{\smallskip}
\end{tabular}

\noindent
Note: We have adopted the molecular abundances for Orion and NGC~7129$-$IRS~2 reported by \citet{Fuente2014}. The listed abundances towards 
NGC~7129$-$IRS~2 correspond to the average value in a region with a radius of D$\sim$0.008~pc centered on the continuum source. For the Orion hot core,
the listed column densities are average values in a region of D$\sim$0.003~pc. For comparison, the spatial resolution of our observations is
HPBW$\sim$(0.005$\times$0.003)~pc. To derive N(SO) we have adopted N(SO)/N(S$^{18}$O)=500 for Orion and NGC~7129$-$IRS~2 . Data for  AFGL 2591 VLA 3 have been taken from \citet{Gieser2019}. In this case,  the synthesized beam is  $\sim$0.007~pc and we have assumed that N(SO)/ N(SO$_2$)$\sim$N($^{33}$SO)/N($^{33}$SO$_2$).
\end{table*}

\section{Summary and conclusions}

We present high angular resolution 1.1mm continuum and spectroscopic ALMA observations of the well-known massive protostellar cluster Mon~R~2$-$IRS~3,
that is composed of two components, IRS~3~A and  IRS~3~B.  The results can be summarized as follows:

\begin{enumerate}

\item Continuum observations show the presence of two components that are separated by $\sim$0.65$"$. These components 
are barely resolved by our 1.1mm observations (HPBW$\sim$ 1.2$"$ $\times$0.9$"$ ). We estimate that IRS~3~A is responsible of $\sim$80 \% of the continuum flux.

\item Spectroscopic observations show that IRS~3~A has the crowded spectra typical of hot cores. In particular, we have detected intense lines of carbon chains such as HCN, H$^{13}$CN, HC$^{15}$N, HNC, and HC$_3$N, the organics CH$_3$OH, HDCO, and H$_2$$^{13}$CO,  the S-bearing species SO, SO$_2$, H$_2$CS, and OCS, and of the COMs, methyl formate (CH$_3$OCHO) and  dimethyl ether (CH$_3$OCH$_3$).

\item Most  species arise in a rotating ring centered on IRS~3~A. In contrast, the emission of COMs peak in the southern part of the NIR reflection nebula. The peak of  the HC$_3$N emission  is displaced by  $\sim$0.5$"$ towards the North of the NIR nebula.

\item The position-velocity diagrams in the direction perpendicular to the NIR nebula reveal that the emission of most molecules comes
from a rotating torus and/or a circumstellar. disk. The  emission of the HCN $\nu_2$=1 3$\rightarrow$2 line remains unresolved and present a linewidth of $\sim$10 km s$^{-1}$. We interpret this emission as coming from the disk wind and/or the emergent outflow. The emission of methyl formate is located in the southern NIR nebula and is more likely associated to the interaction of the bipolar outflow with the molecular cloud.

\item Based on SO$_2$  we derive gas kinetic temperatures of T$_k$$\sim$ 170 K towards the IRS~3~A.  The most abundant S-bearing species is SO$_2$ with an abundance of $\sim$ 1.3$\times$10$^{-7}$, and $\chi$(SO/SO$_2$) $\sim$ 0.29. 

\item At the high temperatures prevailing in hot cores, all the icy mantles of grains are evaporated releasing their content to the gas phase. Following our calculations, SO$_2$ accounts for a significant fraction, $\sim$1\%, of the sulphur budget. The abundance of SO$_2$ is a factor $>$5 higher than those of SO, OCS, and H$_2$CS. This implies that $>$95\% of  the sulphur should be locked in refractories or species such as S$_2$ and sulphur allotropes that cannot be observed.

\end{enumerate}

\section*{Data availability}
The data underlying this article will be shared on reasonable request to the corresponding author.

\section{Acknowledgements}
This paper makes use of the following ALMA data: ADS/JAO. 2016.1.00813.S.
ALMA is a partnership of ESO (representing its member states), NSF (USA) and NINS (Japan), together with NRC (Canada), 
MOST and ASIAA (Taiwan), and KASI (Republic of Korea), in cooperation with the Republic of Chile. The Joint ALMA Observatory 
is operated by ESO, AUI/NRAO and NAOJ. We thank the Spanish MICINN for funding support from AYA2016-75066-C2-2-P 
and PID2019-106235GB-I00. SPTM and acknowledges to the European Union$’$s Horizon 2020 research and innovation program 
for funding support given under grant agreement No 639459 (PROMISE) and Chalmers Gender Initiative for Excellence (Genie).
A.S.M. acknowledges support from the Collaborative Research Centre (SFB) 956 (sub-project A6), funded by 
the Deutsche Forschungsmeneinschaft (DFG) - project 184018867

%%%%%%%%%%%%%%%%%%%%%%%%%%%%%%%%%%%%%%%%%%%%%%%%%%



\bibliographystyle{mnras}
\bibliography{MonR2}

%%%%%%%%%%%%%%%%%%%%%%%%%%%%%%%%%%%%%%%%%%%%%%%%%%

%%%%%%%%%%%%%%%%% APPENDICES %%%%%%%%%%%%%%%%%%%%%

\appendix
\newpage
\section{Supporting material}

\begin{table*}
	\caption{Observations}
	\label{table:observations} 
	\centering                      
	\begin{tabular}{llllll}       
	\hline\hline
Continuum             &  Subbands  (MHz)           &  Synthesized beam   &  PA($^\circ$)   &   $\Delta v$ (km s$^{-1}$)  &  rms (mJy/beam) \\ \hline
1.2mm & 247~GHz+257~GHz    & 1.010$"$ $\times$0.676$"$   &   -64.61 &  4271.07  & 0.87 \\
1.1mm & 262~GHz+272~GHz    & 0.946$"$ $\times$0.634$"$   &   -64.96 &   3248.33 & 0.85 \\ \hline
Spw                &  Freq. range  (MHz)           &  Synthesized beam   &  PA($^\circ$)   &   $\Delta v$ (km s$^{-1}$)  &  rms (K)$^1$ \\ \hline
1  &  255363$-$255490   & 1.23$"$ $\times$0.97$"$ & -65  &    0.25  & 0.145  \\ %HC$^{18}$O$^+$ 
2  &  256185$-$256313        &   1.23$"$ $\times$0.97$"$    & -65   &    0.25  & 0.148  \\ %H29$\alpha$ 
3  &  258127$-$258186        &   1.22$"$ $\times$0.97$"$    & -65   &    0.25  & 0.156  \\ %HC$^{15}$N   
4  &  258197$-$258313  & 1.22$"$ $\times$0.96$"$  & -65  &  0.28  &  0.134 \\ %SO                
5  &  258983$-$259040        &  1.22$"$ $\times$0.97$"$    & -66   &    0.25   & 0.179  \\ %H$^{13}$CN   
6 &  271922$-$272040  &  1.18$"$ $\times$0.94$"$  & -65  &  0.25  &  0.184 \\ %HNC                     
7 &  272835$-$272951  &  1.19$"$ $\times$0.94$"$ & -65  &    0.27  &  0.169 \\ %HC$_3$N              
8 &  273103$-$273219  & 1.17$"$ $\times$0.94$"$  & -65  &  0.25  &  0.211 \\ %l-C$_3$H              
9 &  273550$-$273667  &  1.16$"$ $\times$0.94$"$  & -65  &  0.25  &  0.183 \\ %N$^{15}$NH$^+$  
10  &  274645$-$274782  & 1.16$"$ $\times$0.94$"$  & -65  &  0.25  &  0.178 \\ %c-CH$_3$             
11 &    246171$-$246317 &  1.3$"$ $\times$1.0$"$  & -65  &  0.30  &  0.103  \\ %SO2                     
12 &   247114$-$247260 &  1.3$"$ $\times$1.0$"$  & -65  &  0.30  &  0.105 \\ %C$_3$H               
13  &   248820$-$248944 & 1.3$"$ $\times$1.0$"$  & -65  &  0.25  &  0.102  \\ %C3H2                  
14 &   261946$-$262050  & 1.22$"$ $\times$0.95$"$  & -65  &  0.25  &  0.137  \\ %CCH                     
15 &   261242$-$261300  &   1.23$"$ $\times$0.95$"$  & -65  &  0.25  &  0.180 \\ %HN13C                
16 &    261814$-$261873  &   1.23$"$ $\times$0.95$"$  & -65  &  0.28  &  0.124 \\ %SO7                   
17 &     265702$-$265818  & 1.20$"$ $\times$0.96$"$  & -65  &  0.28  &  0.127 \\ %cHCCCH            
18 &   265828$-$265945   &  1.20$"$ $\times$0.96$"$  & -65  &  0.27  &  0.164 \\ \hline %HCN                    
\hline                                   
\end{tabular}

\noindent
Note:$^1$ The rms of of the spectra towards. IRS~3~A which are shown in Fig.~\ref{fig:espectros1} and .~\ref{fig:espectros2}.
\end{table*}

\begin{table*}
	\caption{Gaussian fits to the continuum emission}
	\label{table:gaussians} 
	\centering                      
	\begin{tabular}{llllll}       
	\hline\hline
RA(J2000)       &   Dec(J2000)      &   Major ($"$)  &  Minor($"$)   &   PA($^\circ$)   &  S$_{1.1mm}$ (mJy/beam) \\ \hline
06:07:47.85     &  $-$06:22:56.27     &    1.09(0.01)    &  0.82(0.01)         &  65.6(0.6)  &  88.5(0.3)  \\
06:07:47.87     &  $-$06:22:55.43     &    1.30(0.01)    &   0.60(0.01)        &  68.1(0.6)  &  25.9(0.3) \\
\hline                                   
\end{tabular}
\end{table*}

\begin{figure*}
    \centering
    \includegraphics[width=0.8\textwidth]{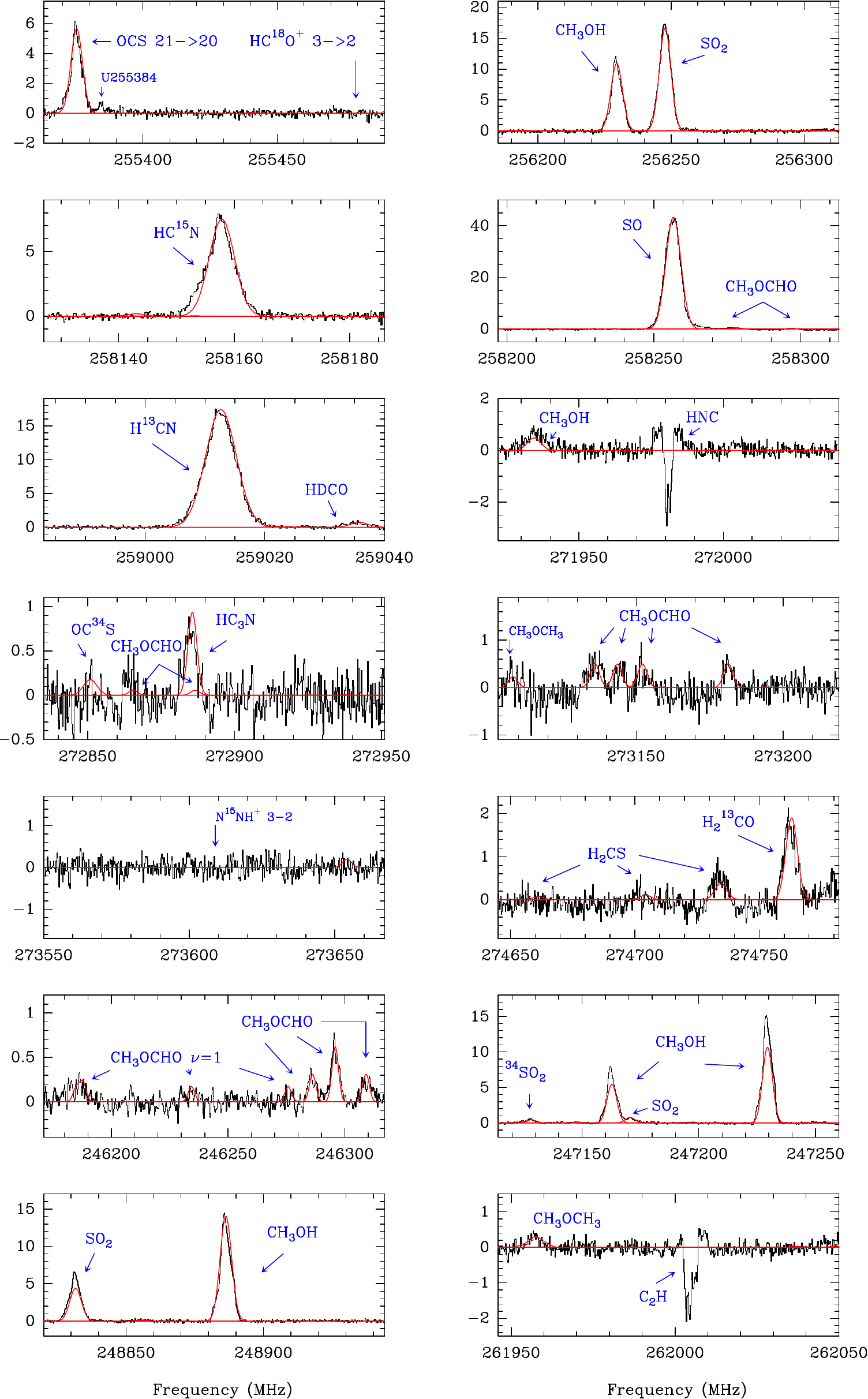}
    \caption{Interferometric spectra towards IRS~3~A. The LTE fitting described in Sect.~\ref{sec:mol} and Table\ref{table:cd} is shown in red.}
    \label{fig:espectros1}
\end{figure*}

\begin{figure*}
    \centering
    \includegraphics[width=0.8\textwidth]{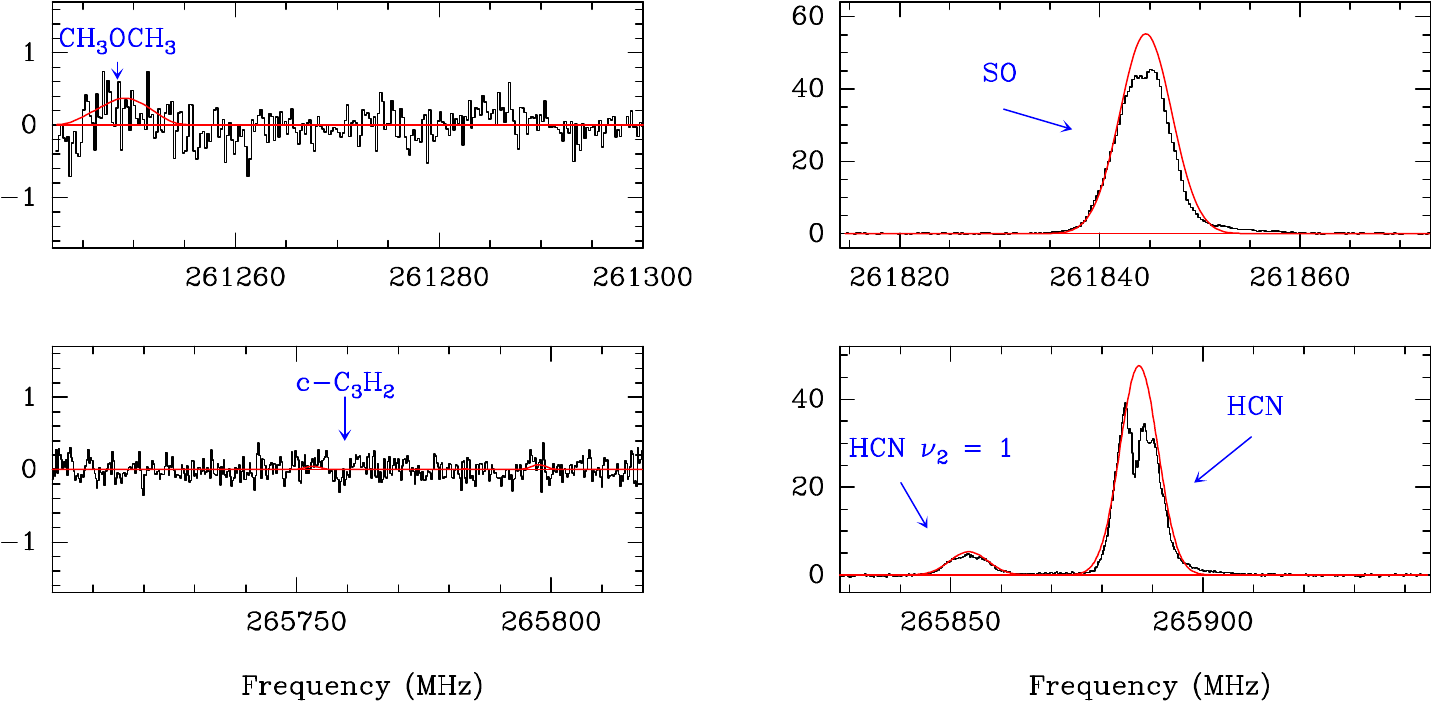}
    \caption{The same as Fig.~\ref{fig:espectros1}.}
    \label{fig:espectros2}
\end{figure*}


%%%%%%%%%%%%%%%%%%%%%%%%%%%%%%%%%%%%%%%%%%%%%%%%%%

% Don't change these lines
\label{lastpage}
\end{document}